\documentclass[final,5p,times,twocolumn,authoryear, ,table,xcdraw]{elsarticle}

\makeatletter
\def\ps@pprintTitle{%
  \let\@oddhead\@empty
  \let\@evenhead\@empty
  \def\@oddfoot{\reset@font\hfil\thepage\hfil}
  \let\@evenfoot\@oddfoot
}
\makeatother

\usepackage{graphicx}%
\usepackage{multirow}%
\usepackage{amsmath,amssymb,amsfonts}%
\usepackage{amsthm}%
\usepackage{mathrsfs}%
\usepackage[title]{appendix}%
\usepackage{xcolor}%
\usepackage{textcomp}%
\usepackage{manyfoot}%
\usepackage{booktabs}%
\usepackage{algorithm}%
\usepackage{algorithmicx}%
\usepackage{algpseudocode}%
\usepackage{listings}%
\usepackage{tabularx}
\usepackage{fancyhdr}
\usepackage{datetime}
\usepackage{multirow}
\usepackage{longtable}
\usepackage{lscape}
\usepackage{supertabular,booktabs}
\usepackage{subfig}

\usepackage{fancyhdr}
\begin{document}

\begin{frontmatter}




\title{\textbf{\centering \Large Behind the Deepfake: 8\% Create; 90\% Concerned\\}\large Surveying public exposure to and perceptions of deepfakes in the UK }

\affiliation[label1]{organization={Public Policy Programme, The Alan Turing Institute},
city={London},
country={UK}}


\affiliation[label2]{organization={Oxford Internet Institute, University of Oxford},
city={Oxford},
country={UK}}

\author[label1]{Tvesha Sippy}
\author[label1]{Florence E. Enock}
\author[label1]{Jonathan Bright}
\author[label1,label2]{Helen Z. Margetts}

\begin{abstract}
 \textit{This article examines public exposure to and perceptions of deepfakes based on insights from a nationally representative survey of 1403 UK adults. The survey is one of the first of its kind since recent improvements in deepfake technology and widespread adoption of political deepfakes. The findings reveal three key }\textit{insights}\textit{. First, on average, 15\% of people report exposure to harmful deepfakes, including deepfake pornography, deepfake frauds/scams and other potentially harmful deepfakes such as those that spread health or religious misinformation or propaganda. In terms of common targets, exposure to deepfakes featuring celebrities was 50.2\%, whereas those featuring politicians was at 34.1\%. Analysing gender and age effects, we find that men and younger people were more likely to report being exposed to deepfakes. And 5.7\% of respondents recall exposure }\textit{to a}\textit{ selection of high profile political deepfakes in the UK. Second, while exposure to harmful deepfakes was relatively low, awareness of and fears about deepfakes were high (and women were significantly more likely to report experiencing such fears than men). As with fears, general concerns about the spread of deepfakes were}\textit{ also}\textit{ high; 90.4\% of the respondents were either very concerned or somewhat concerned about this issue. Specifically, most respondents (at least 91.8\%) were concerned that deepfakes could add to online child sexual abuse material, increase distrust in information and manipulate public opinion. Third, while awareness about deepfakes was high, usage of deepfake}\textit{ tools was relatively low (8\%). And most respondents were not confident about their ability to detect deepfakes. At the same time, most people }\textit{were }\textit{trustful of audiovisual content online. Our work highlights how the problem of deepfakes has become embedded in public consciousness in just a few years; it also highlights the need for media literacy programmes and other policy interventions to address the spread of harmful deepfakes.}  
\end{abstract}




\end{frontmatter}




\section{Introduction \textbf{}}
\label{Introduction}
\subsection{The problem of deepfakes}
The spread of “deepfakes”, a catch-all term referring to audio, image and video content that has been manipulated or created using machine learning methods to alter how a person, object or environment is presented (Centre for Data Ethics and Innovation, 2019), is a source of increasing global concern. Improvements in deepfake creation technology have been significant over the last five years, making them both easier and cheaper to create, as well as being more effective and leading some to present them as one of the most advanced forms of visual disinformation (Weikmann et al., 2024). Their hyper-realism makes it difficult for people to discern them from real media (Busch and Ware, 2023). By enabling alterations to facial features and human voice (Paris and Donovan, 2019), deepfake technologies can be deceptive on multiple sensory levels (Brown et al., 2023), making them powerful drivers of false narratives. Metacognitive experiences associated with audiovisual vis-a-vis textual content may increase deepfakes’ perceived credibility (Vaccari and Chadwick, 2020). Many fear that deepfakes could accelerate the risks of existing misinformation: manipulating public opinion, causing political unrest, influencing voting, encouraging harm based on conspiracies, reinforcing incorrect beliefs, and causing distrust in information (Enock et al., 2024). Even if deepfakes are not more deceptive than false information in textual form (Hameleers et al., 2022), the concern that people may lose faith in all visual material when processing information remains (Weikmann et al., 2024). 

Politically motivated deepfakes are one of the biggest areas of concern. The deepfake audio of US President Joe Biden (Knibbs, 2024), wherein voters from New Hampshire were discouraged from participating in the electoral process, is one prime example. Meanwhile, in the UK, a deepfake of London’s mayor Sadiq Khan was allegedly circulated by a far-right group in 2023; the audio is believed to have caused a stir to local law and order (Spring, 2024). Other politicians including Prime Minister Rishi Sunak and Sir Keir Starmer have also been ‘deepfaked’ in the UK (Good Law Project, 2024). 

Such concerns are not limited to the remit of political deepfakes. Deepfakes are being used to perpetuate financial frauds (Bateman, 2020) and violence against women (Kweilin, 2022). Research suggests that non-consensual pornography constitutes 96\% of deepfake videos found online and women are disproportionately targeted in these videos (Ajder et al., 2019). In a laudable move, the UK government recently criminalised the creation of sexually explicit deepfakes under an amendment to the Criminal Justice Bill. However, experts posit that the legislation falls short, to the extent that it requires proof of perpetrator intent for prosecution (EndViolenceAgainstWomen, 2024). As with sexually explicit deepfakes, legislation for curbing political deepfakes also exist under the Online Safety Act 2023 and the Representation of People Act 1983. However, as experts from the Good Law Project (2024) note, clear regulatory guidance for enforcement is needed. 

However, despite the concern, there is a shortage of recent evidence on the prevalence or impacts of deepfakes, with much existing work (that we review fully below) predating recent improvements in technology. In this article, we aim to fill this gap, by conducting a study that seeks to measure public exposure to deepfakes.  

\subsection{Related Work: Past Measurements of the Spread of Deepfakes }
A characteristic feature of the literature on deepfakes is that it largely focuses on experimental research on the deceptive capacity of deepfakes (Dobber et al., 2021; Hwang et al., 2021; Shin and Lee, 2022) and the detection capacity of humans (Mai et al., 2023; Appel and Prietzel, 2022; Hashmi et al., 2024; Köbis et al., 2021). While such research provides relevant insights for what could potentially work to address the threats posed by deepfakes, we know little about people’s actual exposure to and perceptions of deepfakes outside of controlled research environments.

One of the few primary surveys on public exposure looks at public experiences with deepfake and digitally altered imagery abuse. Flynn et al (2022) found that 14.1\% (n = 864) of respondents experienced one or more forms of abuse: non-consensual creation (11.3\%), non-consensual distribution (10.4\%), and threats to distribute (10.1\%) deepfake and digitally altered images. The survey, conducted in 2019, was an online survey of 6,109 respondents aged 16–64 years across the UK (n = 2,028), New Zealand (n = 2,027), and Australia (n = 2,054). The survey found that both females and males experienced different forms of deepfake image abuse, with males reporting higher overall rates of victimization and perpetration. However, as mentioned above, deepfake technology has advanced considerably since 2019, raising the risk that this evidence is out of date. Besides this survey, there is little primary academic research on public exposure to different types of deepfakes, including political deepfakes, deepfakes used for financial frauds, and non-harmful deepfakes created for educational or entertainment purposes.

It is worth highlighting a considerable amount of grey literature on the subject. For example, a recent YouGov survey (2024) found that two-thirds (67\%) of respondents were concerned about AI-generated misinformation, and (75\%) regarded digitally altered content, such as deepfakes, as a key contributor to the same. The survey, which was nationally representative of the UK, was conducted with over 2000 adults. Similarly, a (2023) cross-country YouGov survey commissioned by Luminate found that a significant majority of citizens – over 70\% in the UK – were worried about the impact of deepfakes on upcoming elections. In 2022, a survey of 16,000 people across eight countries (the US, Canada, Mexico, Germany, Italy, Spain, the UK, and Australia) revealed that 71\% of global respondents were unaware of what a deepfake was, yet most respondents (57\%) were confident in their ability to detect the difference between an authentic video and a deepfake video (iProov, 2022). Furthermore, a little over 60\% of respondents agreed that deepfakes were dangerous. Most respondents were concerned that deepfakes could be used for identity theft, accessing bank accounts, setting up credit cards, and deceiving people into believing something that is not true (iProov, 2022). While these surveys provide relevant information about public perceptions, they adopt a generalised definition of deepfakes, instead of perceptions around specific types of deepfakes. 
Moreover, evidence on public exposure to deepfakes mostly remains restricted to secondary data analytics. For example, in a study on identity fraud, Sumsub (2023) reported a ten fold increase in the number of global deepfakes detected between 2022 and 2023. Similarly, Home Security Heroes (2023), indicated that there were 95,820 deepfake videos in 2023, representing a 550\% increase in volume since 2019. 

Sensity AI’s report (Ajder et al., 2019) provides case studies of how deepfakes were used in politics. The report also indicates that 61\% of non-pornographic deepfakes hosted on YouTube target men, including politicians and corporate figures. Considering non-consensual sexual deepfakes, news reports indicate that apps like Telegram have been used to generate fake nude images of more than 680,000 women. Similarly, TikTok creators are being targeted in non-consensual sexual deepfakes, which is concerning since one-third of TikTok's users are children (Kweilin, 2022). The evidence on deepfake creation is also limited to news and industry reports. Deepfake communities exist on deepfake pornography websites and popular forums like Reddit, 4chan, 8chan, and Voat. In 2019, these communities had approximately 100,000 members across 13 platforms (Ajder et al., 2019). However, Home Security Heroes (2023) estimates a larger ecosystem: with 15 dedicated deepfake creation community websites and forums, and a membership base exceeding 609,464. Home Security Heroes reports that one in three deepfake tools enable users to create deepfake pornography. For example, creating a 60-second deepfake pornographic video, requires only one clear face image, takes less than 25 minutes and costs nothing. The report also found that three-fourths of deepfake pornography users felt no guilt about its usage. 

When considering trust in online content, a YouGov (2024) survey suggests that 81\% of respondents expressed concern about the trustworthiness of online content. While this is an important signal, the survey did not specifically focus on deepfakes, but more broadly on misinformation and AI generated content. Furthermore, there is little research on public perceptions around solutions to this issue. To wit, existing research on deepfakes provides relevant insights on overall trends in the growth of deepfakes, public awareness of and concerns with deepfakes, and interventions that could help detect deepfakes. But this research is largely based on industry reports and experimental research, and does not examine exposure to different types of deepfakes.

\subsection{Research aims} 
By conducting a survey on public exposure to and perceptions of deepfakes, this article makes a unique contribution to the literature. Three other features distinguish this article. First, it measures public exposure to different categories of deepfakes and public concerns about specific consequences of deepfakes, offering more precise insights for policy. While it is well known that politicians and celebrities have been targeted by deepfakes, it is becoming more common for members of the public, particularly teenagers, to be targeted too. Second, it measures public exposure to specific political deepfakes circulated in the UK. Third, it is designed to understand public expectations around actions that social media platforms and other stakeholders should take to tackle this issue, enabling informed decisions on regulations and policy. 

\section{Data and Methods}
\label{Data and Methods}
\subsection{Data collection, ethics, and open science} 
Data collection was conducted online between 31 May and 5 June 2024 and the survey was created and administered using Qualtrics\footnote{www.qualtrics.com}. Participants were recruited through Prolific\footnote{https://www.prolific.com}. The survey was approved by the Ethics Committee at The Alan Turing Institute, UK (approval number: 24062405). Informed consent was obtained at the start of the survey according to approved ethical procedures. The materials and data will be available open access on publication. 

\subsection{Sample} A total of 1403 participants who completed the survey passed standard checks for data quality and were included in the final sample. The sample was designed to be nationally representative of the population of the UK across demographic variables of age, gender and ethnicity (using Prolific’s representative sample tool). To be eligible to participate in the survey, participants had to be 18 or over, fluent in English and a resident in the UK. 

Respondents were aged between 18 and 83 years old, with a mean age of 46.48 (SD = 15.59). A total of 714 participants identified as female (50.9 \%) and 669 as male (47.7 \%). 12 participants identified as non-binary/third gender ($<1\%$), with four selecting ‘prefer not to say’ ($<1\%$), and four selecting ‘prefer to self-describe’ ($<1\%$). The majority were White (1193, 85.0\%), while 114 (8.1\%) were Asian or Asian British, 47 (3.3\%\%) were Black, African, Caribbean, or Black British, and 24 (1.7\%) were mixed, multiple or other ethnicities. 13 ($<1\%$) participants selected ‘any other ethnic group’, whilst 12 ($<1\%$) chose ‘prefer not to say’. Although participants indicated more specific ethnic identities, we have combined them into broader categories to simplify reporting. In terms of education, 781 respondents had degree level qualifications (55.7\%), 215 participants had non-degree level qualifications (vocational or similar) (15.3\%), 403 had no degree level qualifications or equivalent (including completion of secondary school and below) (28.7\%), and 4 participants selected the ‘prefer not to say’ option ($<1\%$). Whilst participants indicated more specific education levels, we combine them into broader categories to simplify reporting. 

\subsection{Survey} \textbf{Demographics and background questions}: We collected standard demographic information about age, gender, ethnicity, education, and political orientation for each participant. Age could be entered as any number with a minimum of 18. For gender, ethnicity, and education level, participants were asked to select the option that they felt best described them from a list of standard predefined categories. For political orientation, participants were asked to select the option that best described where they sat on the political spectrum (more to the left, centre, or more to the right). Participants were also asked how many hours of their personal time they typically spent using the internet per day. All demographic questions, other than their age, provided participants with a ‘prefer not to say’ option (this was not included for age, as being 18 or over was a requirement to participate).

\textbf{\textbf{Awareness:}} We asked participants whether they had heard the term deepfakes before the survey (Yes/No/Not sure). We also asked participants who had previously heard of deepfakes whether they knew what the term meant (Yes/No).

\textbf{\textbf{Exposure:}} To measure exposure to deepfakes, we presented participants with six different types of deepfakes and asked them to indicate if they had personally encountered each one (Yes/No/Not sure). The different deepfake types included: \textit{Deepfakes of politicians; Deepfakes of public figures from the entertainment industry such as actors, social media influencers, musicians; Deepfake pornographic images/videos (nude/sexually explicit images/videos of someone, created without their consent using artificial intelligence); Deepfakes falsifying identities of someone you know for financial frauds/scams (E.g., scammers/fraudsters cloning a family member’s voice to trick you into thinking they are in an emergency situation and need you to send them money); Other deepfakes which you think have the potential to be harmful (E.g., deepfake videos of news reporters/scientists spreading health misinformation or religious leaders spreading propaganda); and Video deepfakes created for entertainment or educational purposes (E.g., deepfake videos of friends with celebrity voice-overs or swapping a friend’s face onto a celebrity/movie character/music video).}

We also asked participants about their exposure to specific deepfakes of politicians that were widely circulated online. The options included: \textit{Audio of Sadiq Khan calling for pro-Palestinian marches; Audio of Sir Lindsay Hoyle badmouthing his aide; Video of a BBC report on a financial scandal involving Rishi Sunak; Audio of Sir Keir Starmer swearing at staffers; Image of Rishi Sunak pouring a sub-standard pint; and Image of Sir Keir Starmer sitting with late sex offender Jimmy Savile.} We also presented participants with placebo deepfakes—deepfake headlines that we invented and that were never actually circulated. These included: \textit{Video of David Cameron withdrawing support for Ukraine, Audio of Sir Keir Starmer admitting to previous cannabis use, Image of Sadiq Khan shaking hands with Jeffrey Epstein, Audio of Liz Truss claiming to have had an affair with Boris Johnson, and Image of Carla Denyer and other members of the green party attending a conference by private jet}. Using the difference between average exposure to deepfakes and placebo deepfakes, we measure true recall of participants to political deepfakes. A similar methodology was followed by Allcott and Gentzkow (2017) in the context of fake news during the 2016 US elections. 

\textbf{\textbf{Fears and Concerns}}: To find out the fears and concerns people have around deepfakes, we asked participants about how fearful they felt about personally becoming a target of harmful deepfakes: \textit{Target of deepfake pornography; Target of deepfake fraud/scam (where your identity is impersonated to trick friends or family into sending money to a fraudster’s account); Target of any other deepfake that could be potentially harmful (E.g., someone creating a deepfake audio/video of you saying something offensive).} Response options were: Not at all fearful /Not very fearful /Somewhat fearful /Very fearful.  Participants were also asked how concerned they felt about the spread of deepfakes on the internet, and the spread of deepfakes having specific consequences: \textit{Contributing to misogyny and online violence against women and girls, for example through the creation of deepfake pornography; Facilitating fraud by impersonating others’ identities (for e.g., audio or visual content pretending to be one of your family members in need, aimed at tricking victims into sending money to the fraudster’s account); Manipulating public opinion through distributing misinformation; Impacting election results or manipulating the political process; Causing general distrust in information; Supporting the narratives of extremist or terrorist groups; Adding to online child sexual abuse material; and Impacting the credibility of evidence during legal proceedings}. Response options for both scales were: Not at all / Not very / Somewhat / Very. 

\textbf{\textbf{Creation}}: To understand access to deepfake creation tools, we asked respondents whether they had ever created a deepfake, either for entertainment, educational or any other purposes, by using a tool for swapping a face with someone else's in a photo or video or generating/editing audio using a voice sample. Response options were: Yes / No / Not sure.

\textbf{\textbf{Detection and Trust}}: We asked respondents whether they thought they would be able to spot if a piece of audio and visual (image and video) content they saw online was a deepfake (Yes / No / Not sure). To understand participants’ perceptions on trust in online audio and visual content, we asked them about the extent to which they trusted that general audio and visual (image and video) content was genuine (i.e., not ‘deepfaked’). Response options were: Do not trust online audio and visual content at all / Do not trust online audio and visual content very much / Trust online audio and visual content somewhat /Trust online audio and visual content very much. Additionally, to understand participants’ views on reliable detection, we asked them to select which stakeholders, if any, they would trust to reliably detect deepfakes. The options were: \textit{Mainstream media; UK government; Social media platforms; Other online providers (e.g., Google); Fact checking organisations; Political parties; AI-based technologies; Scientific experts; Members of the public; Others (free text option); No-one (exclusive); Not sure (exclusive)}.  

\textbf{\textbf{Solutions}}: To understand public perceptions about the solutions to address potentially harmful deepfakes, we asked them two questions. First, what they thought social media platforms should do to tackle the spread of deepfakes which have the potential to be harmful online. Response options were: \textit{Ban or suspend users who distribute deepfakes that may be harmful; Make it easier for people to report deepfakes and request the content to be removed; Make it difficult for users to find specific deepfakes, for example by preventing them from appearing through search terms; Make it harder for people to find tools online which can be used to create harmful deepfakes; Add clear warning labels or watermarks to deepfakes so that origins of the content are clear; Something else (free text option); Nothing - platforms should not do anything to tackle deepfakes (exclusive); None of the above (exclusive); Don’t know (exclusive). }

Second, what other actions respondents thought should be taken to address the spread of harmful deepfakes. Options included: \textit{Bans for applications that allow users to create deepfakes that have the potential to be harmful; Bans for platforms that host deepfakes that have the potential to be harmful, like non-consensual deepfake pornography; Increased training for law enforcement about effectively investigating the creation or distribution of non-consensual deepfakes; Requirements for platforms to systematically report how many potentially harmful deepfakes they host and how they are attempting to combat them; Strict legislation making the creation and distribution of harmful deepfakes illegal;  More education in schools about deepfakes, rules about creating them, and ways to spot them;  Widespread rollouts of media literacy training courses and public awareness campaigns; More funding for research aimed at tackling deepfakes; Regulatory bodies to strictly monitor content violations by BigTech companies and hold them accountable; Public naming and shaming of platforms that fail to deal with harmful deepfakes; Strict biometric/remote identification procedures to protect against deepfake fraud; Something else (free text); Nothing should be done to tackle the spread of online deepfakes (exclusive); None of the above (exclusive); Don’t know (exclusive).} 

\subsection{Procedure}
After participants gave their informed consent to take part in the survey, they responded to the background and demographic questions. Following this, participants responded to questions about their familiarity with deepfakes, before proceeding with  questions about fears and concerns. Next, participants were presented with questions about their exposure to deepfakes and specific political deepfakes.  Post this, participants were asked whether they had created deepfakes, followed by questions about detecting deepfakes, and finally, solutions to address deepfakes. At the end of the questions, participants were given an opportunity to provide feedback in a free text box before continuing to the debrief, and finally completing the submission and being redirected to Prolific for payment. The survey was designed to take approximately 7 minutes to complete, and each participant received £1.05 for their time. 

\begin{figure*}[ht]
    \centering
    \includegraphics[width=1\linewidth]{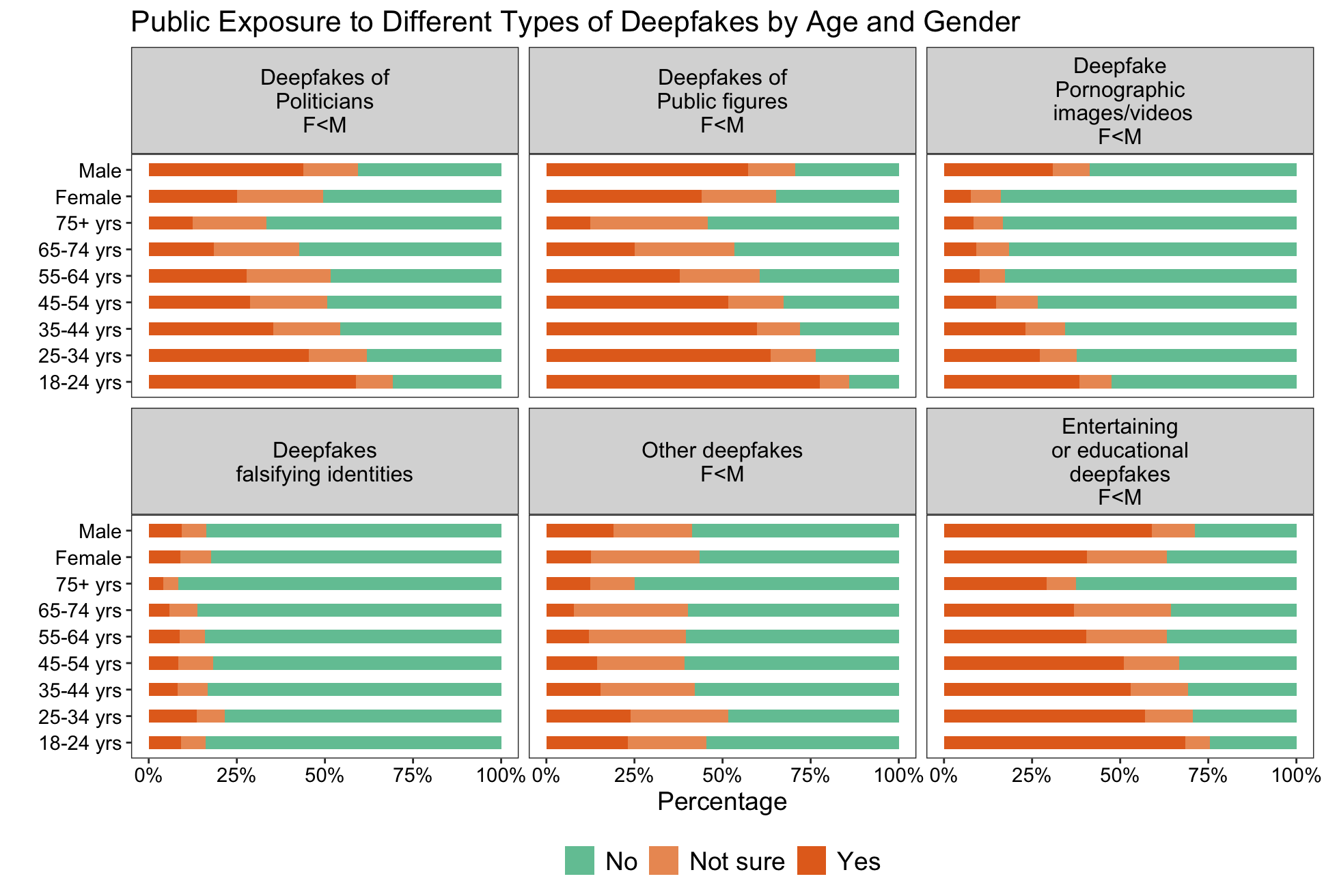}
    \caption{Self reported exposure to different types of deepfakes.  F$<$M indicates that men are significantly more likely to be exposed to deepfakes than women.}
    \label{fig:exposure}
\end{figure*}

\subsection{Data analysis} 
We present descriptive statistics for public exposure to different types of deepfakes, public fears and concerns surrounding deepfakes, and public trust in online content. Additionally, we discuss the top options that respondents select for stakeholders who can be trusted for reliable detection of deepfakes, and solutions that could help address the challenge of deepfakes. As well as describing responses to our key survey questions, we use logistic regressions to test gender and age differences in exposure to and fears of being targeted by deepfakes.

\section{Results}
\label{Key findings}
\subsection{Awareness and creation of deepfakes}
\textbf{Awareness about deepfakes is high: }82.7\% participants had previously heard of the term deepfakes; 12.3\% had not previously heard of deepfakes, and 5.1\% were not sure if they had heard of the term before. Amongst the participants who had previously heard of deepfakes, 91.6\% knew what the term meant. While awareness about deepfakes is high, \textbf{usage of deepfake creation technologies is relatively low}: only 8.1\% respondents indicated that they had created deepfakes for educational, entertainment or other purposes (though 8.1\% of the population creating deepfakes is of course still a significant amount in absolute terms). 

\subsection{Self-reported exposure to deepfakes}
To understand public exposure to deepfakes, we asked respondents whether they had personally encountered six different types of deepfakes. The results show that 50.2\% of people have personally encountered deepfakes of public figures from the entertainment industry such as actors, social media influencers and/or musicians, while exposure to deepfakes of politicians was at 34.1\%. While nearly half of all respondents (49.3\%) had encountered deepfakes created for educational or entertainment purposes, 15.8\% of all respondents were exposed to deepfakes that could be potentially harmful, including deepfakes that spread health or religious misinformation or propaganda. Furthermore, exposure to harmful deepfakes, including non-consensual deepfake pornographic images/videos (18.8\%) and deepfakes that falsify identities for frauds/scams (9\%) were relatively low, though still high enough in absolute terms to indicate a considerable proportion of the population encountering them.

Analysing self-reported exposure to deepfakes by a cross-section of age and gender (see Figure\ref{fig:exposure}), shows that younger participants aged between 18-25 years, report highest exposure to all categories of deepfakes, except deepfakes that falsify identities for frauds/scams and other potentially harmful deepfakes. In these cases, respondents aged between 25-34 years had the highest exposure: 13.6\% and 23.9\%, respectively. 

\begin{figure*}[ht]
    \centering
    \includegraphics[width=1\linewidth]{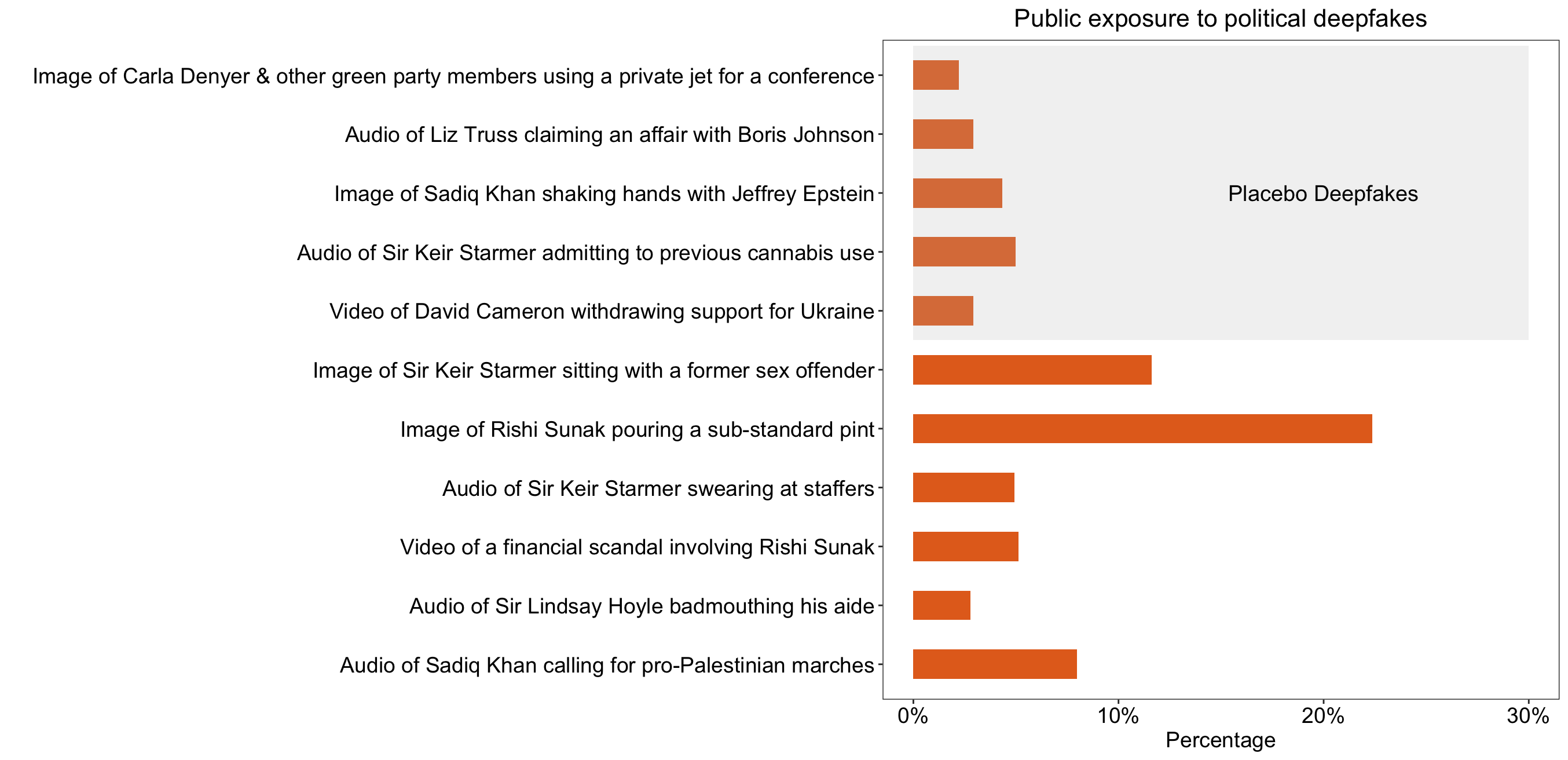}
    \caption{Public exposure to high profile political deepfakes in the UK.}
    \label{fig:exposure_pol}
\end{figure*}

Considering gender, men report higher exposure to deepfakes than women (we note that, as sample sizes are very small, we do not have enough data to present results about those who did not select either male or female as their gender, though this would be important follow up research). To further understand gender and age differences in exposure, we ran a series of logistic regressions, with age and gender as predictor variables. The analysis reveals that across all deepfake categories there was a negative and statistically significant relationship between age and self-reported exposure to deepfakes (p $<0.001$ in all cases except deepfake frauds and scam). And in all cases except fraudulent deepfakes, men were significantly more likely to self-report exposure to deepfakes than women.

\textbf{\textbf{Exposure to specific political deepfakes}: }
As mentioned earlier, governments across the world are particularly worried about the spread of political deepfakes. To better understand public exposure to political deepfakes in the UK, we presented participants with a list of deepfakes and asked them to recall exposure thereto. In addition to deepfakes, we also introduced participants to placebo deepfakes – ie. deepfakes that were not actually circulated, to estimate true recall to political deepfakes. Results show that on average, 9.1\% of respondents reported exposure to deepfakes and 3.5\% reported exposure to placebo deepfakes, as depicted in Figure \ref{fig:exposure_pol}. Using the difference between average exposure to deepfakes and placebo deepfakes as a measure of true recall (following the methodology presented in Allcott and Gentzkow, 2017), we find that on average 5.7\% of respondents in the UK have been exposed to political deepfakes that were circulated online.

\subsection{Concerns and fears about deepfakes}
\begin{figure*}[htbp]
    \centering
    \subfloat{\includegraphics[width=1\linewidth]{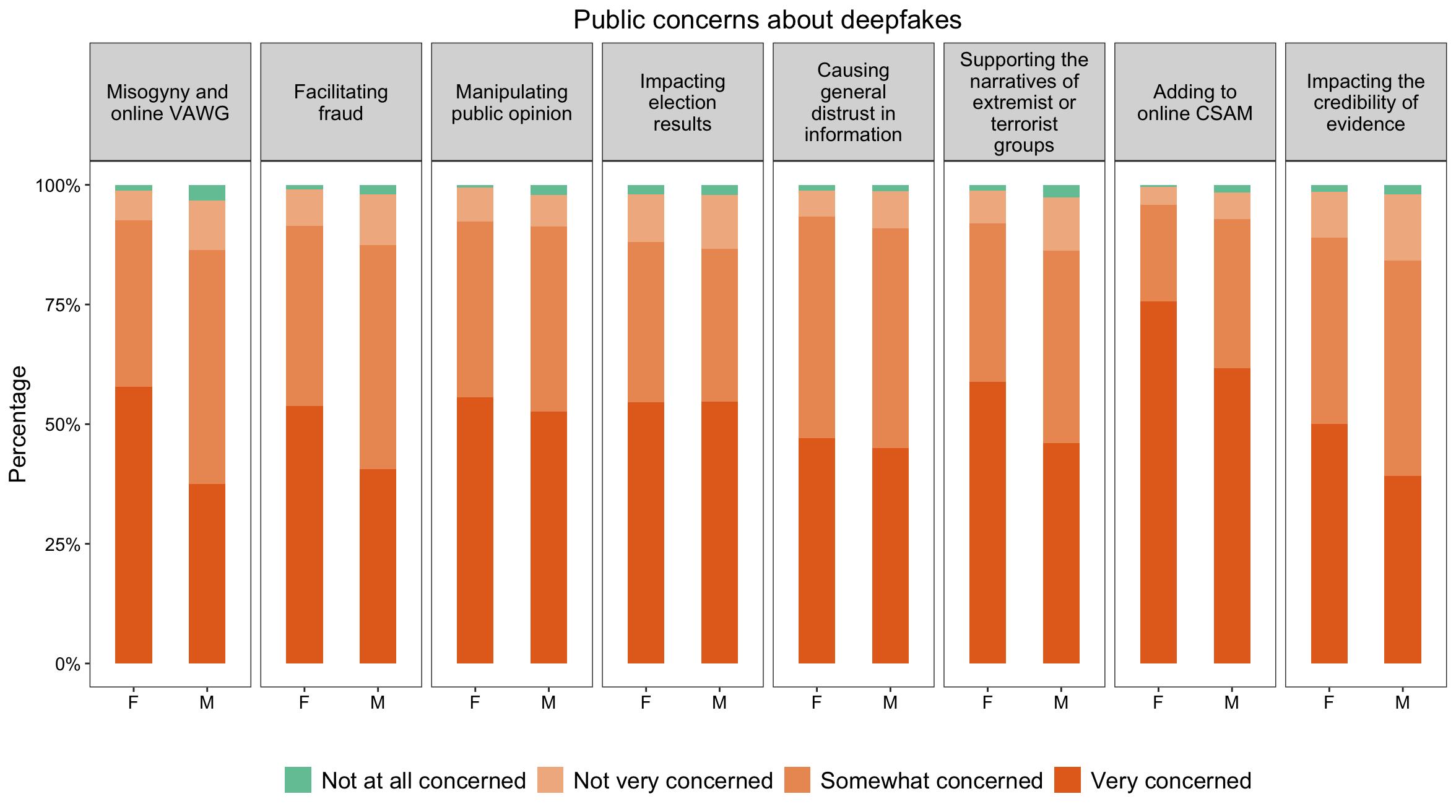}}
    \caption{Public concerns about deepfake consequences}
    \label{fig:concerns}
\subfloat{\includegraphics[width=1\linewidth]{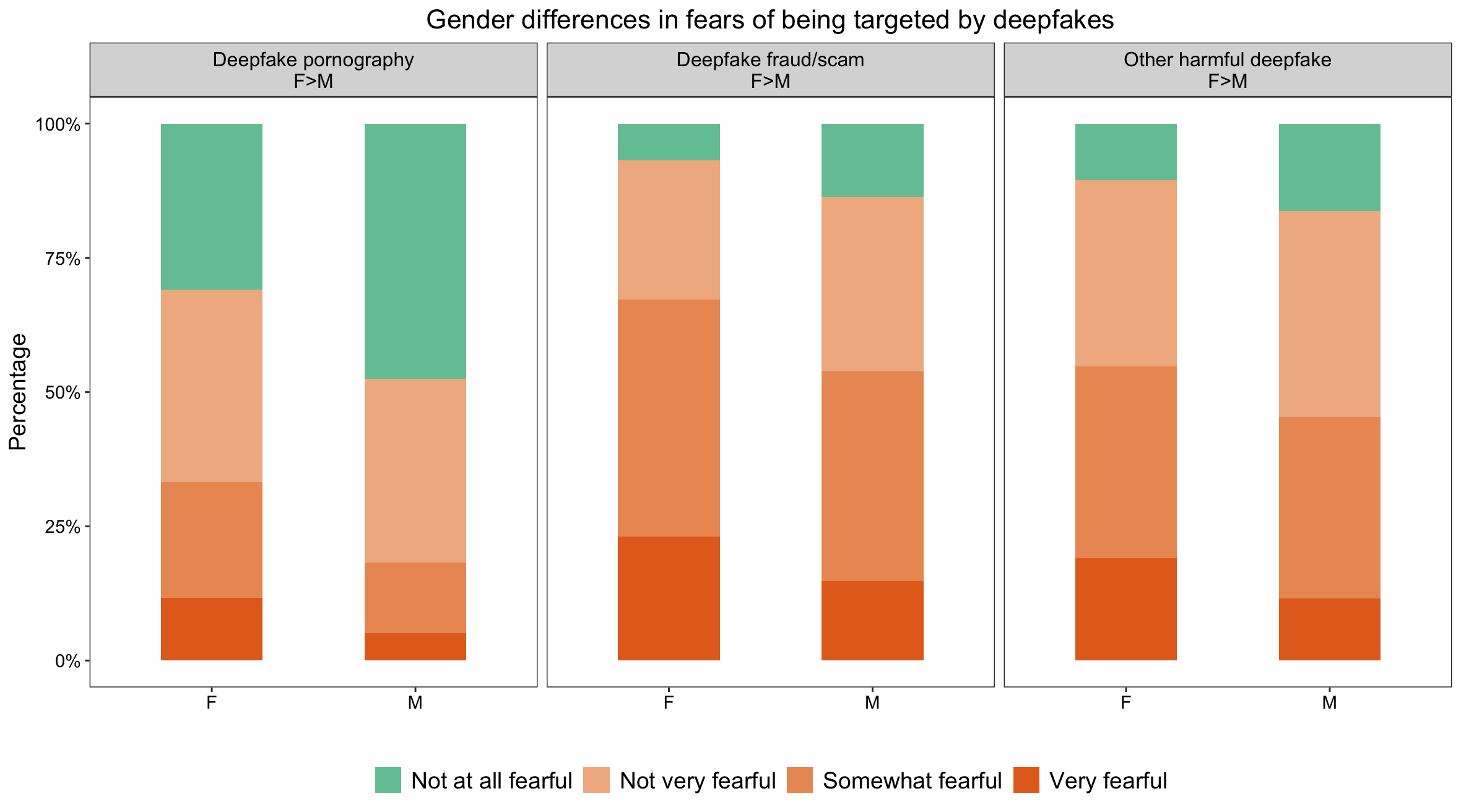}}
\caption{Public fears about becoming a deepfake target.  F$>$M indicates that women are significantly more likely to be fearful of being targeted by deepfakes than men.} 
\label{fig:fears}
\end{figure*}

To understand public concerns about deepfakes, we asked respondents about the extent to which they were concerned about the spread of deepfakes in general, and their concerns about eight potential consequences of deepfakes. Overall, 90.4\% of the respondents were either very concerned or somewhat concerned about the spread of deepfakes. Even concerns across the eight different consequences were high (see Figure \ref{fig:concerns}). The most concerning consequence of deepfakes was in adding to online child sexual abuse material; 94.3\% respondents indicated being very or somewhat concerned about this issue. This was followed by concerns about deepfakes increasing distrust in information (92.1\%) and manipulating public opinion (91.8\%). Respondents also expressed concerns about deepfakes increasing misogyny and online violence against women and girls (89.6\%), facilitating fraud (89.4\%) and supporting the narratives of extremist or terrorist groups (89.1\%). Concerns about deepfakes impacting election results were also high (87.4\%). In fact, 54.7\% people were very concerned about this issue. 

\textbf{\textbf{Fears around deepfakes}:} We asked respondents the extent to which they feared being targeted by harmful deepfakes. The results show (see Figure \ref{fig:fears}) that people were most fearful about being targeted by deepfake frauds/scams. Consider how 60.7\% respondents were either very or somewhat fearful of being targeted by such deepfakes. Half (50.1\%) of all respondents reported being somewhat or very much fearful of becoming targets of other potentially harmful deepfakes, such as deepfakes in which a person is misrepresented as saying something offensive. And a little over one in four respondents (26.1\%) reported being somewhat or very fearful of becoming targets of deepfake pornography. To compare gender differences in fears about becoming a target of each of the 3 harmful deepfake categories,  we created a binary outcome for fear, with `Somewhat' and `Very much' as 1: Fearful, and `Not at all' and `Not very much' as 0: Not fearful. Across all three categories, women were significantly more likely to report experiencing fears than men. Women were 127.4\% (p$<$0.001) more likely to report being fearful of becoming a target of deepfake pornography, 77.9\% (p$<$0.001) more likely to report being fearful of becoming a target of deepfake fraud/scam and  47.7\% (p$<$0.001) more likely to report being fearful of becoming a target of other potentially harmful deepfakes, than men. Table \ref{tab:Fears} in Supplementary Information shows proportions of men and women fearing being targeted by each deepfake type. 

\subsection{Trust in content and deepfake detection}
To understand perceptions of trust in online audio and visual content, we asked respondents to indicate the extent to which they trusted that online content was genuine, ie. not ‘deepfaked’. Results, which were mixed, showed that 30.1\% respondents placed very little trust or no trust at all in the genuineness of online audiovisual content. However, close to two-thirds of respondents (65.1\%) trusted the genuineness of online audiovisual content somewhat, while 4.8\% of respondents trusted the genuineness of online audiovisual content very much. We also asked respondents how confident they felt about spotting deepfakes. While 17.4\% felt confident about spotting deepfakes, 16.5\% were not. Most people (66.1\%) were unsure if they would be able to spot deepfakes.

To understand which stakeholders people trusted to reliably detect deepfakes, we presented them with a list. The top three stakeholders people chose were: fact checking organisations (54.4\%), scientific experts (46.3\%) and AI-based technologies (36.9\%). Table \ref{tab:Stakeholders} in the Supplementary Information shows the proportion of respondents who place trust in each stakeholder. A few respondents, who chose ‘others’, indicated that they would trust the police/MI5, their friends, and other experts including people with prior knowledge of image manipulation technology, tech-savvy hackers, photo and video editors, specific Discord or YouTube accounts that specialise in deepfakes and visual effects, photographic analysis experts and private tech companies. 

We also asked respondents how they would respond if they suspected that a piece of content was `deepfaked'. A majority of respondents indicated that they would: not engage with the content (54.9\%) and report it to the social media platform they discovered it on (52.1\%). The next most popular responses were researching the content’s origin (31.6\%) and using a fact-checking tool (29.5\%). Table \ref{tab:Response} in the Supplementary Information shows the proportion of respondents for each response type. A few respondents who selected ‘other’ in response to this question reported that they would add a community note on X (formerly Twitter) or report to Action Fraud UK.

\subsection{Solutions}
To understand what people thought should be done about the problem of deepfakes, we asked respondents two sets of questions. 

The first concerned the actions that social media platforms should take to address the issue. Overall, participants supported social media platforms taking action to tackle the harmful content they host; less than 1\% chose ‘Nothing - platforms should not do anything to tackle deepfakes’. Most respondents (87.3\%) expressed that platforms should ban or suspend users who create harmful deepfakes, and 82.4\% of participants noted that social media platforms should make it easier for people to report harmful deepfakes and request for content removal.  Table \ref{tab:Platform_Actions} in the Supplementary Information shows the overall proportion of participants favouring each action. 

A few respondents, who selected ‘something else’, suggested that platforms should educate people, use tools provided by AI providers to discern synthetic media and enforce content moderation, while others expressed imposing bans/fines for platforms that host deepfakes. 

Second, we asked participants to indicate other solutions that could help address the deepfake challenge (see Table \ref{tab:Other_Actions} in the Supplementary Information). The top two most commonly chosen solutions were strict legislation making the creation and distribution of harmful deepfakes illegal (72.1\%) and bans for platforms that host deepfakes that have the potential to be harmful, like non-consensual deepfake pornography (71.8\%). Besides this, 69.3\% and 69.1\% respondents chose the following two actions, respectively: more education in schools about deepfakes, and increased training for law enforcement about effectively investigating the creation or distribution of non-consensual deepfakes. 

Suggestions of respondents who selected ‘something else’ includes fines for platforms, a database of known deepfakes, disclaimers on posts featuring politicians and royals, making it illegal to create deepfakes, and requiring people to take personal responsibility and be more aware about these issues. 
 
\section{Discussion and conclusion}
To understand public exposure to and perceptions of deepfakes amongst the UK public, we asked a nationally representative sample of 1403 people whether they had encountered different categories of deepfakes online, what their fears and concerns surrounding deepfakes were, and which actions they thought social media platforms and the government should take to address the spread of potentially harmful deepfakes online. 

Results suggest that most respondents were aware about deepfakes and knew what the term meant. This stands in stark contrasts to the (2022) iProov survey, which found that 71\% of global respondents, including those from the UK, were unaware of what a deepfake was. Between 2022 and 2024, high profile deepfakes of politicians and celebrities such as Taylor Swift, may have brought deepfakes to the centre of the public discourse, accounting for increased awareness. Self-reported exposure to deepfakes was high (49.3\%) in cases of non-harmful deepfakes including those created for educational or entertainment purposes. In terms of targets, exposure to deepfakes featuring public figures from the entertainment industry (50.2\%) exceeded exposure to deepfakes featuring politicians (34.1\%). Exposure to more disconcerting deepfakes, including non-consensual deepfake pornography, fraud-related deepfakes, and deepfakes used to perpetuate misinformation or propaganda was relatively low. In terms of demographics, men and younger participants reported highest exposure to deepfakes. 

Despite relatively low exposure to harmful deepfakes, concerns about deepfakes were consistently high. The most concerning consequences were that deepfakes could add to online child sexual abuse material, increase distrust in information, and manipulate public opinion–with at least 91.8\% of respondents expressing being either ‘very or somewhat’ concerned about each of these issues. In terms of deepfakes impacting election results, 87.4\% respondents were concerned. When read with a (2023) cross-country survey poll commissioned by Luminate, which found that over 70\% of citizens in the UK were worried about the impact of deepfakes on upcoming elections – it appears that these concerns may be growing. However, it is important to note that only 5.7\% respondents recall being exposed to specific political deepfakes circulated within the UK.  The effects of specific political deepfakes on public perceptions remains to be researched. In 2019, the Responsible Technology Adoption Unit (previously known as the Centre for Data Ethics and Innovation) noted that only a few political deepfakes had emerged on social media platforms, and they were yet to see a “convincing deepfake of a politician that could distort public discourse.” In 2024, weeks before the elections in the UK, these observations appear to still hold true. 

In terms of public fears, our survey finds that people were most fearful (60.7\%) about being targeted by deepfake frauds/scams, followed by other potentially harmful deepfakes, such as deepfakes in which a person is misrepresented as saying something offensive (50.1\%) and deepfake pornography (26.1\%). And across all three categories, gender differences in fears revealed that women were significantly more likely to experience fears than men. Despite relatively low exposure to these harmful deepfakes, public fears remain high. Furthermore, close to two-thirds of all respondents were unsure if they would be able to spot deepfakes.  

We also find nuances in people’s perceptions of trust. While most respondents were concerned about deepfakes causing general distrust in information, close to 70\% report trusting online audiovisual content. What this suggests is people’s attitudes about trust varies when thought through the lens of the individual vis-a-vis society. Future research must examine the effects of exposure to deepfakes on individual and group-level trust in audiovisual content. It is also important to note here that not all deepfakes are harmful. Deepfakes may be used for educational or entertainment purposes, as our survey also reveals. The study did not cover people’s attitudes towards such beneficial cases, although this may also shape overall trust in audiovisual content. 

In terms of policy-relevant insights, the survey finds that people trust fact checking organisations (54.4\%), scientific experts (46.3\%) and AI-based technologies (36.9\%) to reliably detect deepfakes. This aligns with recent research, which suggests that AI models are better at detecting deepfakes compared to humans (Hashmi et al., 2024). Considering possible solutions, the most common actions respondents selected include banning or suspending users who created harmful deepfakes and requiring platforms to make it easier for people to report harmful deepfakes and request for content removal. The latter is a particularly important civil remedy that should be institutionalised, since not all people targeted by harmful deepfakes would prefer to pursue criminal proceedings. 

Other actions suggested by respondents include strict legislation making the creation and distribution of harmful deepfakes illegal, bans for platforms that host deepfakes that have the potential to be harmful such as non-consensual deepfake pornography, more education in schools about deepfakes and increased training for law enforcement about effectively investigating the creation and/or distribution of non-consensual deepfakes. While the needle on legislation has certainly moved forward, much more needs to be done for effective enforcement. Busch and Ware (2023) note that governments should train the public in identifying signs of deepfakes and corroborating content from different sources. “Pre-bunking” measures may help the public better respond to deepfakes when they encounter them. 

There are several limitations to this research. First, we note that we asked about exposure to different types of deepfakes. However, there may be overlaps in the different categories we asked about. For example, there may be cases where a politician is featured in a non-consensual sexually explicit deepfake. Because of this, we cannot separate out these intersections between targets and nature of deepfakes. Second, due to limitations posed by a small sample size, we were unable to assess whether ethnic minority groups were more exposed to certain types of deepfakes or experienced greater fear. Furthermore, we were unable to analyse the experiences of gender minorities due to limitations of sample size. Future research should examine such intersections and diverse voices. It must also examine experiences of teenagers and young adults, who are known to be targets of non-consensual sexually explicit deepfakes. Third, self-reported exposure may be under- or over- reported as people may be unsure about their exposure due to difficulties in identifying deepfakes. 

Taken together, our findings show that there is increasing awareness of and concerns about deepfakes amongst the UK public. At the same time, relatively high trust in the genuineness of online content shows that people might need to be more sceptical of claims made in audiovisual content online – and media literacy programmes could help here. People also expect stronger action from both platforms and governments to address the challenges posed by harmful deepfakes.
 
\section*{Corresponding author:}
\label{Author contact}
Tvesha Sippy, tsippy@turing.ac.uk 


\section*{Acknowledgements}
This work was supported by the Ecosystem Leadership Award under the EPSRC Grant EPX03870X1 and The Alan Turing Institute. 

\section*{References}
Ajder, H., Patrini, G., Cavalli, F., \& Cullen, L. (2019). The State of Deepfakes: Landscapes, Threat and Impact. Deeptrace Labs

Allcott, H., \& Gentzkow, M. (2017). Social Media and Fake News in the 2016 Election. Journal of Economic Perspectives, 31(2), 211–236. 

Appel, M., \& Prietzel, F. (2022). The detection of political deepfakes. Journal of Computer-Mediated Communication, 27(4).

Bateman, J. (2020). Deepfakes and Synthetic Media in the Financial System: Assessing Threat Scenarios. Carnegie Endowment for International Peace

Brown, J. G., J. N. Bailenson, \& J. Hancock. 2023. “Misinformation in Virtual Reality.” Journal of Online Trust and Safety 1:1–31.

Busch, E., \& Ware, J. (2023). The Weaponisation of Deepfakes. Digital Deception by the Far-Right. Policy Brief. International Centre for Counter-Terrorism

Dobber, T., Metoui, N., Trilling, D., Helberger, N., \& de Vreese, C. (2021). Do (Microtargeted) Deepfakes Have Real Effects on Political Attitudes? The International Journal of Press/Politics, 26(1), 69–91. 

EndViolenceAgainstWomen. (2024). Government criminalises creation of deepfakes, but with a major loophole. End Violence Against Women. 

Enock, F.E., Bright, J., Stevens, F., Johansson, P., \& Margetts, H. Z. (2024). How do people protect themselves against online misinformation? Attitudes, experiences and uptake of interventions amongst the UK adult population. The Alan Turing Institute.

Flynn, A., Powell, A., Scott, A. J., \& Cama, E. (2022). Deepfakes and Digitally Altered Imagery Abuse: A Cross-Country Exploration of an Emerging form of Image-Based Sexual Abuse. The British Journal of Criminology, 62(6), 1341–1358. 

Good Law Project. (2024). Clamping down on political deepfakes. 

Hameleers, M., Van Der Meer, T. G. L. A., \& Dobber, T. (2022). You Won’t Believe What They Just Said! The Effects of Political Deepfakes Embedded as Vox Populi on Social Media. Social Media + Society, 8(3)

Hashmi, A., Shahzad, S. A., Lin, C.-W., Tsao, Y., \& Wang, H.-M. (2024). Unmasking Illusions: Understanding Human Perception of Audiovisual Deepfakes (arXiv:2405.04097). arXiv. 

Home Security Heroes. (2023). State Of Deepfakes: Realities, Threats, And Impact.

Hwang, Y., Ryu, J.Y., and Jeong, S.-H. (2021). Effects of disinformation using deepfake: the protective effect of media literacy education. Cyberpsychol. Behav. Social Netw. 24, 188–193.

iProov. (2022). Deepfake Statistics \& Solutions | Protect Against Deepfakes. 

Knibbs, K. (2024). Researchers Say the Deepfake Biden Robocall Was Likely Made With Tools From AI Startup ElevenLabs. WIRED. 

Köbis, N. C., Doležalová, B., \& Soraperra, I. (2021). Fooled twice: People cannot detect deepfakes but think they can. iScience, 24(11), 103364.

Kweilin, T. L. (2022). Deepfakes and Domestic Violence: Perpetrating Intimate Partner Abuse Using Video Technology. 

Luminate. (2023). Bots versus ballots: Europeans fear AI threat to elections and lack of control over personal data. 

Mai, K. T., Bray, S., Davies, T., \& Griffin, L. D. (2023). Warning: Humans cannot reliably detect speech deepfakes. PLOS ONE, 18(8), e0285333. 

Paris, B., \& Donovan, J. (2019). Deep Fakes and Cheap Fakes: The Manipulation of Audio and Visual Evidence. Data\&Society. 

Shin, S. Y., \& Lee, J. (2022). The Effect of Deepfake Video on News Credibility and Corrective Influence of Cost-Based Knowledge about Deepfakes. Digital Journalism, 10(3), 412–432. 

Spring, M. (2024). Sadiq Khan says fake AI audio of him nearly led to serious disorder. BBC News. 

Sumsub. (2023). Sumsub Identity Fraud Report 2023.

The Centre for Data Ethics and Innovation. (2019). Independent Report: Deepfakes and Audiovisual Disinformation. GOV.UK.

Vaccari, C., \& Chadwick, A. (2020). Deepfakes and Disinformation: Exploring the Impact of Synthetic Political Video on Deception, Uncertainty, and Trust in News. Social Media + Society, 6(1)

Weikmann, T., Greber, H., \& Nikolaou, A. (2024). After Deception: How Falling for a Deepfake Affects the Way We See, Hear, and Experience Media. The International Journal of Press/Politics, 19401612241233539.

YouGov. (2024). Can you trust your social media feed? UK public concerned about AI content and misinformation

\section{Supplementary Information}
\newpage

\label{SI}

\begin{table}
\begin{tabular}{>{\raggedright\arraybackslash}p{0.3\linewidth}>{\raggedright\arraybackslash}p{0.2\linewidth}>{\raggedright\arraybackslash}p{0.15\linewidth}>{\raggedright\arraybackslash}p{0.15\linewidth}}
\textbf{Deepfake type}& \textbf{Fear}& \textbf{Male} & \textbf{Female} \\ \hline
Deepfake fraud/scam& Not fearful& 46.2\%& 32.77\%\\
 & Fearful& 53.8\%& 67.23\% \\ \hline
Deepfake pornography& Not fearful& 81.8\%& 66.81\%\\
 & Fearful& 18.2\%& 33.19\% \\ \hline
Other potentially harmful deepfake& Not fearful& 54.7\%& 45.24\%\\
 & Fearful& 45.3\%& 54.76\%\\\end{tabular}
 \caption{Level of fear reported by men and women about becoming a deepfake target. Women are significantly more fearful than men at the p$<$.001 level.}
    \label{tab:Fears}
\end{table}
\begin{table}
\begin{tabular}{>{\raggedright\arraybackslash}p{0.6\linewidth}l}
\textbf{Stakeholder}& \textbf{Response (\%})\\ \hline
Fact checking organisations& 54.4\%
\\ \hline
Scientific experts& 46.3\%
\\ \hline
AI-based technologies& 36.9\%
\\ \hline
UK government& 25.2\%
\\ \hline
Mainstream media& 22.2\%
\\ \hline
Other online providers (e.g., Google)& 14.3\%
\\ \hline
Social media platforms& 14.1\%
\\ \hline
Members of the public& 13.3\%
\\ \hline
No-one& 12.6\%
\\ \hline
Not sure& 10.5\%
\\ \hline
Political parties& 5.0\%
\\ \hline
Other, please specify& 1.8\%\\\end{tabular}
 \caption{Stakeholders who people trust to reliably detect deepfakes}
    \label{tab:Stakeholders}
\end{table}
\begin{table}
\begin{tabular}{>{\raggedright\arraybackslash}p{0.6\linewidth}l}
\textbf{Response to deepfakes}& \textbf{Response (\%})\\ \hline
Decide not to engage with it (e.g., through liking/sharing it)& 54.9\%
\\ \hline
Report it to the platform or site you saw it on& 52.1\%
\\ \hline
Attempt to research its origin (e.g., using a search engine)& 31.6\%
\\ \hline
Use a fact-checking tool& 29.5\%
\\ \hline
Ask the opinion of someone else, like a friend or family member& 27.0\%
\\ \hline
Comment on the content to warn others& 20.7\%
\\ \hline
Report it to an independent regulator (Ofcom)& 12.1\%
\\ \hline
Nothing& 9.3\%
\\ \hline
Report it to the police& 7.8\%
\\ \hline
Not sure& 4.5\%
\\ \hline
Something else (please specify)& 1.1\%\\\end{tabular}
 \caption{Response to potential deepfakes}
    \label{tab:Response}
\end{table}
\begin{table}
\begin{tabular}{>{\raggedright\arraybackslash}p{0.5\linewidth}l}
\textbf{Platform-based solutions}& \textbf{Response (\%})\\ \hline
Ban or suspend users who distribute deepfakes that may be harmful& 87.3\%
\\ \hline
Make it easier for people to report deepfakes and request the content to be removed& 82.4\%
\\ \hline
Add warning labels or watermarks to deepfakes so that origins of the content are clear& 76.6\%
\\ \hline
Make it harder for people to find tools which can be used to create harmful deepfakes& 56.8\%
\\ \hline
Make it difficult for users to find specific deepfakes, for example by preventing them from appearing through search terms& 55.5\%
\\ \hline
Don't know& 1.9\%
\\ \hline
Something else (please specify)& 1.7\%
\\ \hline
Nothing - platforms should not do anything to tackle deepfakes& 0.3\%
\\ \hline
None of the above& 0.1\%\\\end{tabular}
 \caption{Platform-based solutions for responding to the challenge of deepfakes}
    \label{tab:Platform_Actions}
\end{table}
\begin{table}
\begin{tabular}{>{\raggedright\arraybackslash}p{0.6\linewidth}l}
\textbf{Other solutions}& \textbf{Response (\%})\\ \hline
Strict legislation making the creation and distribution of harmful deepfakes illegal& 72.1\%
\\ \hline
Ban platforms that host deepfakes that have the potential to be harmful, like non-consensual deepfake pornography& 71.8\%
\\ \hline
More education in schools about deepfakes, rules about creating them, and ways to spot them& 69.3\%
\\ \hline
Increased training for law enforcement about investigating the creation or distribution of non-consensual deepfakes& 69.1\%
\\ \hline
Requirements for platforms to report how many potentially harmful deepfakes they host and how they are attempting to combat them& 62.1\%
\\ \hline
Ban applications that allow users to create deepfakes that have the potential to be harmful& 60.8\%
\\ \hline
Public naming of platforms that fail to deal with harmful deepfakes& 59.6\%
\\ \hline
Widespread rollouts of media literacy training courses and public awareness campaigns& 51.1\%
\\ \hline
More funding for research aimed at tackling deepfakes& 45.0\%
\\ \hline
Don't know& 2.4\%
\\ \hline
Something else (please specify)& 1.1\%
\\ \hline
None of the above& 0.4\%\\ \hline
Nothing should be done to tackle the spread of online deepfakes& 0.3\%\\\end{tabular}
 \caption{Other actions for responding to the challenge of deepfakes}
    \label{tab:Other_Actions}
\end{table}

\end{document}